\documentclass[amsmath,notitlepage,twocolumn,pra]{revtex4-1}
\RequirePackage{varioref}
\RequirePackage{amssymb}
\RequirePackage{bm}
\RequirePackage{graphicx}
\RequirePackage{color}

\newcommand{\com}[1]{}

\newcommand{\mathcommand}[3][0]{\newcommand{#2}[#1]{\ensuremath{#3}}}
\renewcommand{\vec}[1]{\ensuremath{\text{\textbf{#1}}}} % Simply a bold charactor.

\newcommand{\isotope}[2]{\ensuremath{^{#2}\text{#1}}}
\newcommand{\Si}{\isotope{Si}{29}}
\newcommand{\Ge}{\isotope{Ge}{73}}

\newcommand{\be}{\begin{equation}}
\newcommand{\ee}{\end{equation}}
\newcommand{\ts}[2]{{#1}_{\textnormal{#2}}} %A math symbol with a text subscript
\newcommand{\tsc}[2]{{#1}_{\textsc{\lowercase{#2}}}} %A math symbol with a text subscript

\newcommand{\Tr}[2][]{\text{Tr}_\text{#1}\left\{#2\right\}}

\mathcommand[1]{\smallket}{|#1\rangle}
\mathcommand[1]{\smallbra}{\langle #1|}
\mathcommand[1]{\bigket}{\bigl|#1\bigr\rangle}
\mathcommand[1]{\bigbra}{\bigl\langle #1\bigr|}
\mathcommand[1]{\biggket}{\biggl|#1\biggr\rangle}
\mathcommand[1]{\biggbra}{\biggl\langle #1\biggr|}
\mathcommand[1]{\ket}{\left| #1 \right\rangle}  %a Ket
\mathcommand[1]{\lket}{\bigl| #1 \bigr)}  %a Liouville Ket
\mathcommand[1]{\bra}{\left\langle #1 \right|}  %a Bra
\mathcommand[1]{\lbra}{\bigl( #1 \bigr|}  %a Bra
\newcommand{\ketbra}[2]{\left| #1 \right\rangle\!\!\left\langle #2
	\right|} %an outer product
\newcommand{\braket}[2]{\left\langle #1 |#2\right\rangle} %an inner product
\mathcommand[1]{\bbra}{\biggl\langle #1 \biggr|}
\mathcommand[1]{\bket}{\biggl| #1 \biggr\rangle}
\newcommand{\refeq}[1]{Eq.~\eqref{#1}}

\begin{document}

\renewcommand\graphicspath[1]{#1}   % remove subdir for arXiv {./magnetic_noise_figures/#1}
\title{Magnetic Gradient Fluctuations from Quadrupolar \Ge\ in \\ Si/SiGe Exchange-Only Qubits}
\author{J.~Kerckhoff}
\author{B.~Sun}
\author{B.~H. Fong}
\author{C.~Jones}
\author{A.~A.~Kiselev}
\author{D.~W.~Barnes}
\author{R.~S.~Noah}
\author{E.~Acuna}
\author{M.~Akmal}
\author{S.~D.~Ha}
\author{J.~A.~Wright}
\author{B.~J.~Thomas}
\author{C.~A.~C.~Jackson}
\author{L.~F.~Edge}
\author{K.~Eng}
\author{R.~S.~Ross}
\author{T.~D. Ladd}
\affiliation{HRL Laboratories, LLC,
3011 Malibu Canyon Rd., Malibu, CA 90265}
\date{\today}

\begin{abstract}
We study the time-fluctuating magnetic gradient noise mechanisms in pairs of Si/SiGe quantum dots using exchange echo noise spectroscopy.  
We find through a combination of spectral inversion and correspondence to theoretical modeling that quadrupolar precession of the \Ge\ nuclei play a key role in the spin-echo decay time $T_2$, with a characteristic dependence on magnetic field and the width of the Si quantum well.   
The \Ge\ noise peaks appear at the fundamental and first harmonic of the \Ge\ Larmor resonance, superimposed over $1/f$ noise due to \Si\ dipole-dipole dynamics, and are dependent on material epitaxy and applied magnetic field.  
These results may inform the needs of dynamical decoupling when using Si/SiGe quantum dots as qubits in quantum information processing devices.
\end{abstract}
\maketitle

\begin{figure*}[th!]
	\includegraphics[width=\textwidth]{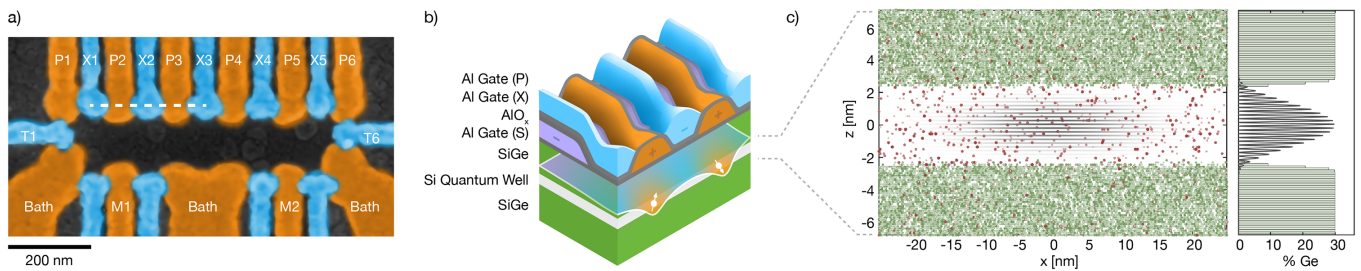}
	\caption{a) False-color scanning electron micrograph of a six dot device.
		A 2DEG is accumulated under the bath gates (Bath), and loaded through the tunnel barriers from by gates (T).
		The plunger gates (P) form the quantum dots and exchange gates (X) controls the exchange interaction between neighboring dots.
		The measure dot (M) acts as an electrometer, sensing the charge state in each dot through capacitive shift.  This work focuses on data from double quantum dots on devices featuring arrays of up to six dots, as highlighted.
		b) Cartoon of the Si/SiGe heterostructure, as labelled; gate stack, color-matched to the white-dashed-line-indicated region of (a), and a depiction of a double-dot potential, which is depressed due to P gates and lifted due to X gates; electrons with spin are depicted as white.
		c) Illustration of the Si/SiGe barrier and electron wavefunction of a single dot with population of spin-carrying nuclear spins.  The plot places a green circle for each \Ge\ nucleus and a red circle for each \Si\ nucleus in a 50$\times$50$\times$14~nm slab of material (dot size and shade indicate distance into the out-of-page $y$ dimension).   The dots are randomly placed at each silicon lattice site with probability 800~ppm for \Si\ in both well and barrier and 7.76\% for \Ge\, multiplied by the alloy content.   The sidewall content of \Si\ is enriched in this figure, but is natural in the actual devices.  The alloy content for Ge is shown as horizontal bars on the right, for a profile with 1 atomic layer of smearing at the Si/SiGe interface. This Ge percentage is used to determine the vertical wavefunction $|\psi(z)|^2$, shown in black and featuring valley oscillations with phase determined by Ge placement.  A 30-nm-diameter Gaussian wavefunction diameter is assumed in the $x$ and $y$ directions.}
	\label{fig:Dev}
\end{figure*}

\section{Introduction}
\label{sec:intro}

Qubits made of silicon are promising candidates for quantum computation for a variety of reasons~\cite{Ladd2018}.
One of these reasons is the long coherence times observed for spins in bulk silicon crystals, going back many decades~\cite{Gordon1958}.
These long spin coherence times are due to the quiet magnetic environment of the surrounding crystal and allow spin qubits in silicon to hold quantum information for timescales long relative to typical control-pulse timescales.
This advantage is especially evident when comparing to spin-qubits based on III-V semiconductors such as GaAs, in which every substrate nucleus has a nuclear magnetic moment.
Critically, the best coherence times are observed when silicon is isotopically enhanced; that is, when the silicon source has a depleted population of the spin-1/2 \Si\ isotope, leaving behind only the natural spin-0 isotopes of \isotope{Si}{28} and \isotope{Si}{30}~\cite{Tyryshkin2012,Muhonen2014,Veldhorst2014,Eng2015,Veldhorst2015,Jock2018,Yoneda2018,Andrews2019,struck_low-frequency_2020}.

Silicon spin qubits may host electrons in a variety of ways, including binding them to single donors~\cite{Muhonen2014}, trapping them at a metal-oxide-semiconductor (MOS) interface~\cite{Veldhorst2014,Veldhorst2015,Jock2018}, or trapping them in a strained silicon quantum well in a Si/SiGe heterostructure~\cite{Kawakami2014,Zajac2015,Eng2015,Reed2016,Zajac2018,Watson2018,Yoneda2018,Andrews2019,struck_low-frequency_2020}.
Here, we focus on the SiGe-based quantum well approach. Quantum dots in SiGe are lithographically defined and benefit from the low disorder of the Si-SiGe interface, advantages relative to donor bound spins or MOS, but suffer from the disadvantages of lower valley splittings and reduced compatibility with many commercial complementary MOS (CMOS) fabrication processes~\cite{Ladd2018}.

One of the key questions about SiGe-based dots not encountered for MOS or donor qubits is whether and how the only stable nonzero-spin isotope of Ge, \Ge, impacts magnetic noise~\cite{witzel_nuclear_2012}.
In the present work, we examine devices in which the \Si\ content is depleted from the natural levels of 4.7\% to 800~ppm in the Si well, and address the critical question of whether the remaining magnetic noise will be dominated by these residual \Si\ nuclei, by naturally-abundant \Ge\ nuclei, or by other noise sources.  
In particular, we focus on magnetic limits to coherence in spin-echo experiments (i.e. $T_2$ times) and find through both modeling and dependence on material epitaxy that these coherence limits are driven by the distribution of \Ge\ nuclei in the device when applied magnetic fields are low.  
This question informs what engineering or material modifications may be needed in the future to allow high fidelity quantum gates on qubits based on these materials.

Magnetic noise in silicon qubits has been studied previously in different types of devices and qubit encodings, which we now briefly review.
Single qubit gates may employ exchange only~\cite{DiVincenzo2000,Eng2015,Andrews2019}, electron spin resonance (ESR)~\cite{Muhonen2014,Veldhorst2014}, or electron dipole spin resonance (EDSR)~\cite{Kawakami2014,Zajac2018,Yoneda2018,struck_low-frequency_2020}; two qubit gates may either involve exchange only~\cite{DiVincenzo2000} or a combination of exchange and radio-frequency control~\cite{Veldhorst2015,Zajac2018,Watson2018,fogarty_integrated_2018,zhao_single-spin_2019}.
In all cases, fluctuations of local magnetic fields from \Si\ nuclei or other sources will impact gate fidelity, determining the eventual overhead of quantum error correction for allowing quantum information processing~\cite{Veldhorst2016b,Jones2018}.
For isotopically enhanced donor and MOS-based single-spin qubits at high magnetic field, long static dephasing times, exceeding 100~$\mu$s, have been observed, and spin-echo studies indicate low-frequency magnetic noise dominated by fluctuations in the applied magnetic field ~\cite{Muhonen2014,Veldhorst2014,madzik_controllable_2020}.
Higher-frequency ($>$100 Hz) magnetic noise in these devices may be dominated by the effect of charge motion on the position of the electron, which translates to magnetic noise due the hyperfine coupling to the \isotope{P}{31} nucleus in the case of the donor or to a spin-orbit-induced Stark shift in the case of MOS. In SiGe, existing studies to date employ large magnetic field gradients induced by a microferromagnet, allowing single-spin control via EDSR ~\cite{Kawakami2014,Kawakami2016,Yoneda2018,Zajac2018,struck_low-frequency_2020}.
In these studies, in both isotopically natural and isotopically enhanced samples, magnetic noise appears to be dominated by electric-field noise transduced by the deliberately applied gradient.

Here, we study the nuclear environment of SiGe qubits without introducing any deliberate magnetic gradients, and in a low-field environment where spin-orbit and paramagnetic or diamagnetic gradients from local screening effects are minimized.
This is the magnetic environment of exchange only qubits~\cite{DiVincenzo2000,Eng2015,Andrews2019}, and in this regard, we examine a simpler magnetic environment than previous studies.
In this simple environment, higher frequency magnetic fluctuations appear to be entirely dominated by residual nuclear magnetic moments, in contrast to previous studies.  Using exchange echo noise spectroscopy, we find a rich noise spectrum including components originating from quadrupolar \Ge\ nuclear moments and provide a detailed model for these effects.

\section{The $\text{SiGe}$ Quantum Dot Device}
\label{sec:device}
We study accumulation mode devices in this manuscript predominantly employing a gate stack similar to that described in Ref.~\onlinecite{Borselli2015}, except with aluminum gates to control quantum dot loading and a large aluminum screening gate to prevent electrons from accumulating under the gate leads, leading overall to a gate-stack very similar to that in Ref.~\onlinecite{Zajac2015}.  There are several variations in the gate stack across different samples, but the aspects varied are not expected to cause significant variation in $T_2^*$ or $T_2$ as is our focus here.

All devices were fabricated on $[100]$-oriented Si$_{0.7}$Ge$_{0.3}$/Si/Si$_{0.7}$Ge$_{0.3}$ buried channel heterostructures grown on thick graded buffers using a variety of methods as described in Ref.~\onlinecite{Deelman2016}.  
Silicon well widths varied from 3 to 10 nm, set between a 170 nm SiGe buffer below and a 60 nm Si$_{0.7}$Ge$_{0.3}$ barrier above, typically also including a 1.5 nm Si cap. 
An SEM image of an example six-dot device is shown in Fig.~\ref{fig:Dev}a, and crossection schematic in Fig.~\ref{fig:Dev}b.  
For most samples, the Si wells were isotopically enriched to 800 ppm \isotope{Si}{29}, while the Si$_{0.7}$Ge$_{0.3}$ uses a natural abundance of \isotope{Ge}{73}.  
The samples in this paper use natural abundance Si in the SiGe sidewalls, although we will argue in Sec.~\ref{sec:t2} that the silicon isotopic abundance of the barriers plays little role in qubit behavior due to the dominant contribution of isotopically natural barrier \Ge.  

Some samples have been characterized by atomic probe tomography, as discussed in Ref.~\cite{Dyck2017}, but in most cases the detailed shape of the Si/SiGe barrier is not known.  
Figure~\ref{fig:Dev}c shows a model for the barrier and wavefunction for a 5~nm well, discussed further in Sec.~\ref{sec:t2}, with an illustration of the density of \Si\ and \Ge\ nuclei.  
This figure illustrates that the wavefunction appreciably overlaps hundreds of \Si\ nuclei in the well and hundreds of \Ge\ nuclei at the Si/SiGe barrier; a more quantitative evaluation of hyperfine overlap is discussed in Sec.~\ref{sec:t2}.  In contrast with the actual devices, the figure depicts enriched levels of \Si\ in the sidewalls to show the location of the \Ge\ nuclei more clearly.  One open question for the present study is whether magnetic gradient noise spectroscopy may help determine the degree of nuclear overlap, and hence details of the Si/SiGe barrier.
This is especially plausible if the effect of $^{73}$Ge can be isolated from that of $^{29}$Si, which we will see is enabled by both the use of isotopically enriched Si in the quantum well and the use of noise spectroscopy.

While the SiGe-Si-SiGe heterostructure confines electrons to the Si quantum well along the growth direction,  voltages applied to the plunger (P) gates create an attractive potential which accumulates and confines single electron spins inside the quantum well plane. 
The tunnel barriers (T) control the loading rate of single electrons from the accumulated baths. 
Applying a voltage to the exchange (X) gates changes the potential barrier between dots and is used to control the exchange interaction between neighboring dot pairs. 
The device shown in Fig.~\ref{fig:Dev} features six dots, or two triple-quantum-dot exchange-only qubits~\cite{DiVincenzo2000,Eng2015,Andrews2019}, however all experiments described in this paper utilize only two dots at a time.  
The experiments described here may be performed on any dot-pair of the device, with similar results.  
In addition to the dots formed under the P gates, two additional dots are formed under the two measure (M) dot gates, which are used as electrometers to sense the occupation of each dot through capacitive shifts. 

All experiments use a standard energy-selective initialization procedure \cite{Petta2005,Eng2015,Reed2016} to prepare a singlet state, $\ket{S}\equiv[\ket{\uparrow\downarrow}-\ket{\downarrow\uparrow}]/\sqrt{2}$, which has total spin amplitude $\mathcal{S}=0$, in the first two dots.
Spin-state measurement is performed using a Pauli-spin blockade as sensed by the M-dot electrometers which can distinguish between singlet states, with $\mathcal{S}=0$, and triplet states, $\mathcal{S}=1$, but cannot discriminate between the triplet projection substates, $m = -1,0,1$.  Coherent spin swaps are performed by ``symmetric'' control of the exchange \cite{Reed2016}. 
Calibration of such pulses use all three dots operating as a single qubit and the calibration procedure is described in Ref.~\onlinecite{Andrews2019}.  All three operations are depicted schematically in Fig.~\ref{fig:CPn_Schem}a.

While the spin singlet and $m=0$ triplet states are insensitive to global magnetic fields, local magnetic fields such as hyperfine-coupled nuclear spins create fluctuations of magnetic gradients between quantum dots, driving transitions between singlet and triplet states.  We can effectively homogenize the magnetic fluctuations by coherently swapping the spins, so that each spin is equally subjected to the same averaged local magnetic environment.
This is the basis for exchange-based spin echo. Exchange echo sequences can use an arbitrary number of exchange pulses and in the following section we will calculate the effect of such a pulse sequence on a double-dot initialized to a spin singlet. 

\section{Exchange-based Singlet-Triplet Magnetic Noise Spectroscopy}
\label{sec:singlet-triplet}

Noise spectroscopy is a well-studied suite of techniques, especially for qubits, for deducing noise spectral characteristics from an ensemble of measurements using different noise-compensating sequences~\cite{Alvarez2011,Bylander2011,norris_optimally_2018}.  Versions of noise spectroscopy have been applied to a variety of spin-qubit systems, and in particular to singlet-triplet qubits in GaAs~\cite{botzem_quadrupolar_2016,malinowski_notch_2017}.  We provide here a brief theoretical summary of singlet-triplet magnetic noise spectroscopy as pertinent to the experiments we describe in this paper; we detail the theory further in Appendix~\ref{app:FF}.  For this, we will neglect charge noise, initialization errors, and measurement errors and assume a spin singlet is perfectly prepared and measured, and spin-swaps occur perfectly and instantaneously.  While this approximation will inevitably fail for many-pulse experiments, we will discuss no more than 10-pulse experiments in the present work and we find no deleterious effects of charge noise on our ability to extract magnetic noise phenomena.

We first note that for singlet-triplet qubits, we are immune to global magnetic fields, since the singlet state $\ket{S}$ is an invariant eigenstate relative to any global magnetic field, including dynamically fluctuating global magnetic fields.  However, if the magnetic field on dot 1 and the magnetic field on dot 2 vary in magnitude, direction, or both, the singlet state will evolve into orthogonal triplet states, which is the effect we seek to characterize.  The time-dependent, noisy Hamiltonian of the system is simply
\begin{multline}
	H(t)=J(t)\vec{S}_1\cdot\vec{S}_2
	\\-g\ts\mu{B}[B_0^z(S_1^z+S_2^z)
	+\delta\vec{B}_1(t)\cdot\vec{S}_1
	+\delta\vec{B}_2(t)\cdot\vec{S}_2],
\end{multline}
where $\vec{S}_k$ is the vector spin-operator for electron $k$ with $z$-component $S_k^z$, $g\approx 2$ is the electron's $g$-factor, $\ts\mu{B}$ the Bohr magneton, $B_0^z$ is a global applied magnetic field taken to be in the $z$ direction, and $J(t)$ describes voltage-modulated exchange energy.  For the present work, we will consider only instantaneous $\pi$ pulses, and hence take $J(t)$ as $\sum_j
\hbar\pi\delta(t-t_j)$ for pulses arriving at prescribed times $t_j$.  
Differences in the local magnetic field variations at each dot, $\delta\vec{B}_k(t)$, represent the noisy fluctuation whose spectral character we seek to deduce via control of the spin-swap-pulse arrival times $t_j$. 
We can summarize the free evolution for a pair of spins via a unitary operator
\be
U(t) = e^{-i\vec{b}_1(t)\cdot
	\vec{S}_1}e^{-i\vec{b}_2(t)\cdot\vec{S}_2},
\ee
in which $\vec{b}_k(t)$ represents the effective axis and angle about which spin $k$ rotates, which includes a combination of noisy magnetic vector fields and spin-swaps.  In general, the relationship between $\vec{b}_k$ and $\vec{B}_k$ is complex; only if all effective magnetic field vectors are parallel and no exchange is applied can we make the simplifying assumption $b^z_k=\int_0^t dt g\ts\mu{B} (B_0^z+\delta B_k^z(t))/\hbar$.

In all experiments, we initialize $\ket{S}$ and measure the return probability to $\ket{S}$, and average over a time-ensemble of measurements.  Appealing to an ergodic ensemble average, which we notate $\langle\cdot\rangle$ and discuss further at the end of this section, the singlet probability measured in each experiment can then be approximately summarized (see Appendix~\ref{app:FF}) as
\begin{multline}
	\label{PS}
	P_S(t) \approx \bigl\langle |\!\bra{S}U(t)\ket{S}\!|^2 \bigr\rangle
	\\
	\approx
	\frac{1}{2}+\frac{1}{2}\cos\biggl(
	\bigl\langle|\vec{b}_1-\vec{b}_2|t\bigr\rangle\biggr)\exp\biggl(-\frac{ 
		\sigma_{12}^2(t)}{2}\biggr).
\end{multline}
Our focus in the present work is
the second moment $\sigma_{12}^2(t)$,  which captures dynamic fluctuations around random mean magnetic field gradients, which may occur at a variety of timescales.  We decompose this second moment using a filter function formalism as
\be
\label{FFdef}
\frac{\sigma_{12}^2(t)}{2}
= \frac{g^2\ts\mu{B}^2}{2\hbar^2} \sum_{k=1}^2 \int_0^\infty df S_k(f) F(f,t),
\ee
where $S_k(f)$ the power spectral density for each component of the magnetic field noise in dot $k$.  
The filter function $F(f,t)$ depends on the experiment being performed.  
Experimentally, we only have access to a total noise power $\sum_k S_k(f)$, however we derive filter functions $F(f,t)$ scaled for one dot in Eq.~(\ref{FFdef}) and assume independent, identical noise distributions across the pair.  
Details of the calculation method for any number of spin-swaps are discussed in Appendix~\ref{app:FF}.  There, we show that the filter function can be decomposed into three terms; a ``central lobe" which we notate as $F_0(f,t)$ and replicas of that central lobe at positive and negative electron Larmor frequencies $\omega_0=g\ts\mu{B}B^z_0/\hbar.$  
These side-bands appear whenever the noisy magnetic field gradients in the system have a ``transverse" component, meaning that their vector directions include components orthogonal to the applied magnetic field.
This in turn means that the experiment must somehow drive the system to compensate for the electron Zeeman energy, rendering these side-lobes negligible at sufficiently high magnetic fields.
In the experiments described here, a field of 10~mT is sufficiently large to completely suppress these sidelobes.

\begin{figure*}
	\includegraphics[width=\textwidth]{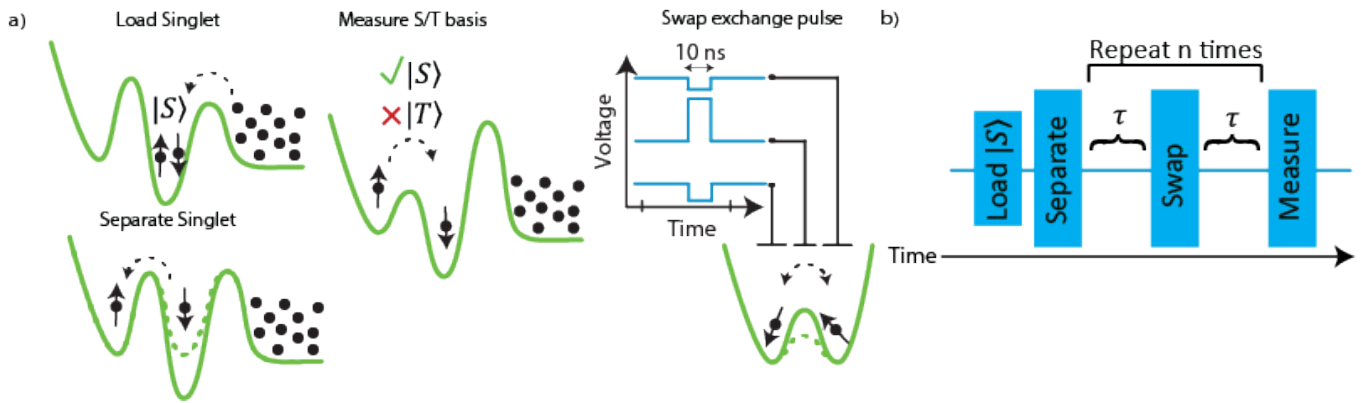}
	\caption{a) Schematic representation of double dot intialization, single/triplet measurement, and pulsed exchange swaps.  b) Representation of the experimental protocol for Hahn echo (n=1) and Carr-Purcell-n experiments in our double dot system.}
	\label{fig:CPn_Schem}
\end{figure*}

\begin{figure}
	\includegraphics[width=\columnwidth]{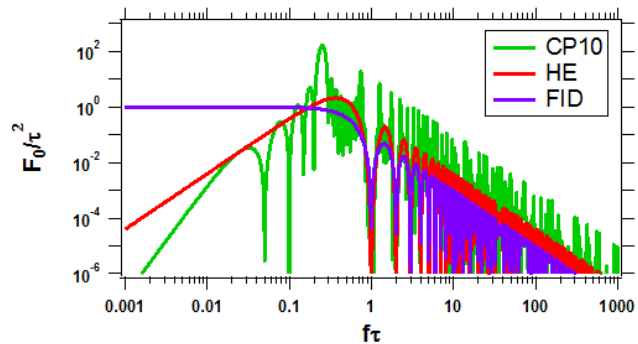}
	\caption{Center lobe filter functions for Carr-Purcell-10 (CP10), Hahn Echo (HE), and Free Induction Decay (FID) experiments.}
	\label{fig:Filter_Funcs}
\end{figure}

In some experiments we may be sensitive to a first moment $\langle|\vec{b}_1-\vec{b}_2|\bigr\rangle$; this will depend on any static magnetic field gradients across the pair of dots.  Static magnetic field gradients may arise from the screening of the applied magnetic field by the superconducting aluminum gates, at fields beneath the critical field, or by spin-orbit effects at higher magnetic fields.  We observe both phenomena in our devices, which we elaborate in Appendix~\ref{app:static_gradients}.  

A primary role of exchange echo experiments is to cancel the low-frequency noise leading to a finite first-moment by undoing static or nearly static phase evolution.
This is accomplished using one or more calibrated spin swaps via voltage-induced exchange in between equal-time durations of evolution in the (1,1) charge state, which alternate which spin sees which dot's magnetic field.  
If using only a single swap, we refer to the experiment as a Hahn echo (HE) experiment, and its central-lobe filter function following Appendix~\ref{app:FF} is
\begin{equation}
\label{HEFF}
\text{HE}  : F_0(f,\tau) = \frac{\sin^2(2\pi f \tau)\tan^2(\pi f \tau)}{(\pi f)^2}.
\end{equation}
Here $\tau$ is the amount of time of free evolution before and after the spin swap.
If some number $n$ swaps are used, with $n>1$, we refer to the experiment as a Carr-Purcell-$n$ (CP$n$) experiment~\footnote{Note that in a Bloch sphere defined by the singlet and $m_z=0$ triplet state $\ket{T_0}=[\ket{\uparrow\downarrow}+\ket{\downarrow\uparrow}]/\sqrt{2}$, the spins are prepared along
	the same axis as the rotation caused by exchange $\pi$-pulses, and therefore this latter experiment may be considered a Carr-Purcell-Meiboom-Gill (CPMG) sequence when the number of
	pulses is even, but we refer to these as CP$n$ nonetheless.}.
For even $n$, the filter function is altered to
\begin{equation}
\label{CPnFF}
\text{CP}n : F_0(f,\tau) = \frac{4\sec^2(2\pi f \tau)\sin^4(\pi f \tau)\sin^2(2n\pi f \tau)}{(\pi f)^2}.
\end{equation}
In this case, $\tau$ is the amount of free evolution time before first and after the last swap; between swaps spins evolve for the time $2\tau$.  
The protocol for these experiments is depicted in Fig.~\ref{fig:CPn_Schem}b.

Both Eq.~(\ref{HEFF}) and Eq.~(\ref{CPnFF}) vanish at $f=0$, indicating their capability to decouple low frequencies~\cite{Cywinsky2008}.   Intuitively, we may think of these experiments as ``homogenizing" the magnetic field by swapping the two spins sufficiently rapidly between the dots, converting static local gradient fields into global fields, to which singlet-triplet subspaces are immune.  The associated filter functions are depicted in Fig.~\ref{fig:Filter_Funcs}, which shows $F_0(f,\tau)$ for CP10 and HE.  
These two filter functions act as narrow- and broad-band filter functions, respectively (plus additional peaks at odd integer multiples).  

The key concept of noise spectroscopy is to scan these band-pass filters across the noise spectrum by varying $\tau$ in a series of experiments.   
The resulting observed decay may then be inverted to estimate the underlying spectrum.  Our approach for this inversion, following Refs.~\onlinecite{Alvarez2011} and \onlinecite{Bylander2011}, is to first correct the data using the signal at $\tau\rightarrow 0$ and $\tau\rightarrow\infty$ to extract a decay function $\exp(-\chi(t))$, where $t=2n\tau$, from the singlet probability of form $[1+\exp(-\chi(t))]/2$.  
We then treat the noise under study as roughly constant within the bandwidth of the main lobe of the filter function, which for CP$n$ extends from $(1-2/n)/(4\tau)$ to $(1+2/n)/(4\tau)$, centered at $f=1/(4\tau)$.  
The integral under this lobe is $0.732n\tau$, giving a summed spectral density across both dots of 
\begin{equation}
\sum_k S_k(f)\approx \frac{8f\chi(n/2f)}{0.732n}.
\label{inversion_definition}
\end{equation}
While more sophisticated noise spectroscopy inversion techniques certainly exist~\cite{norris_optimally_2018}, we find this simple inversion method sufficient for deducing the sources and magnitudes of underlying fluctuating gradient fields.

We may also analyze the decay by extracting $T_2$, which we define as the total sequence time (i.e., $2n\tau$) at which coherence has decayed to its $1/e$ point as $\tau$ is increased and $n$ fixed in HE and CP$n$ experiments.  This $1/e$ definition has limited predictive value, especially since different noise spectra can lead to a significant variation of decay shapes.   For example, if integrated against a simple $1/f$ spectrum, both filter functions for CP10 and HE result in a decay going as $\exp[-(t/T_2)^2]$ (i.e., one that is Gaussian in time).  If integrated against white noise, exponential decay would result, and significantly different $T_2$ values could be chosen for comparable noise powers.  Our chosen definition of $T_2$ therefore is insufficient to predict dephasing at different timescales; the underlying deduced spectrum is necessary to project the impact of dephasing.  

Finally, also included in Fig.~\ref{fig:Filter_Funcs}  is the filter function for a so-called Free Induction Decay (FID) experiment, which has no spin swap pulses and is thus sensitive to noise components at low frequencies, as discussed more in Appendix~\ref{app:static_gradients}.  The $1/e$ point of this decay provides the experimental definition of $T_2^*$, but we caution that $T_2^*$ depends on slow fluctuations that are more challenging to spectroscopically invert.

We now briefly address ergodicity: in the expressions above, we have used an average over an ensemble of independent dot-pairs, but in our experiments we study only a single dot-pair evolving over the long timescale of a time-averaged ensemble of experiments. The ergodic assumption is violated if the averaging timescale is too short relative to the lowest noise frequency under study.
This discrepancy is particularly acute for power-law noise spectra, $S(f)\propto 1/f^\alpha$, which diverge as $f\rightarrow 0$, and experiments highly sensitive to low-frequency noise, in particular FID, are particularly sensitive to insufficient averaging.  
The noise sources we will study in this paper, however, all appear to achieve an ``ergodic" limit at some finite averaging time, indicative of a low-frequency deviation from true $1/f^\alpha$ behavior, and we present data throughout with sufficient averaging to have seen this limit reached.  See Ref.~\cite{Eng2015} as an example characterization of the ergodic limit in a double-dot magnetic noise experiment; similar characterizations have been performed for the measurements in this study.

\section{Fluctuating Magnetic Field Gradient Measurements}
\label{sec:hifreq}

\begin{figure}
	\includegraphics[width=\columnwidth]{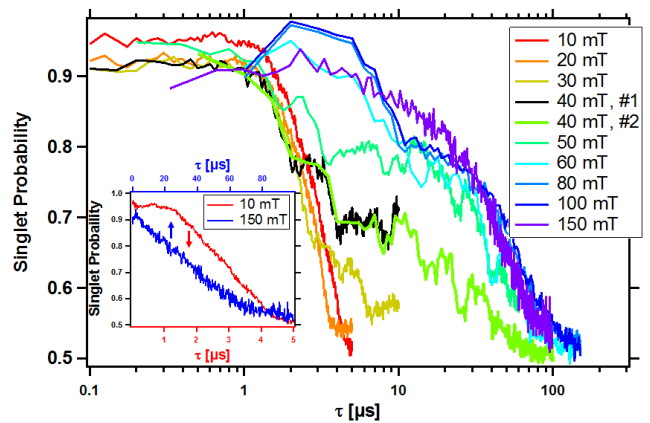}
	\caption{CP10 echo decay curves vs. $\tau$ measured with external magnetic fields ranging from 10 mT to 150 mT.  Singlet survival times increase dramatically as the field is increased and complex decay curves appear at intermediate fields.  The green and black curves at 40~mT are two separate measurements, showing that the observed structure is reproducible.  Inset: 10 mT and 150 mT data on linear $\tau$ scales.}
	\label{fig:CP10vB}
\end{figure}

\begin{figure}
	\includegraphics[width=\columnwidth]{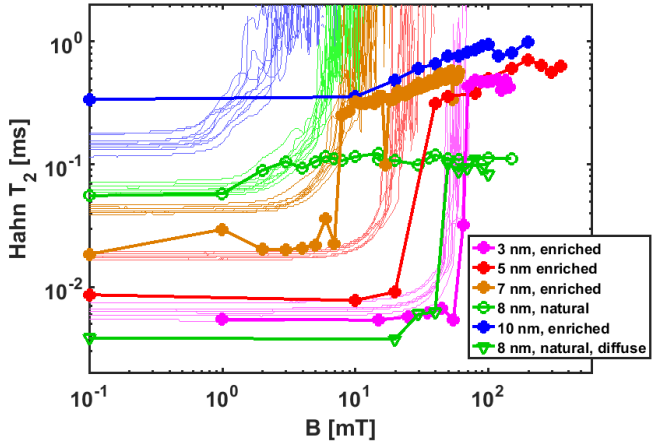}
	%Fig_T2Hahn_vs_B_vs_Well_exp.png}}
	\caption{Hahn Echo $T_2$ as determined by $1/e$ decay time, versus well width and applied magnetic field.  Thick lines and symbols are experimental measurements across six devices; thin lines are results of simulations for 10 devices at each well width with varying random \Ge\ placement.}
	\label{fig:T2vWidthExp}
\end{figure}

In this section we summarize singlet-triplet noise spectroscopy measurements from HE and CP$n$ experiments and a variety of devices to help illuminate the magnetic dynamics of \isotope{Ge}{73}.  

\subsection{Time-domain Decay}

Figure~\ref{fig:CP10vB} shows the decay observed from CP10 echo experiments performed with an applied magnetic field ranging from 10~mT to 150~mT (always applied in the in-plane direction in this work, approximately along the $[1\bar{1}0]$ crystal axis), from a device with a 5 nm Si quantum well enriched to 800 ppm $^{29}$Si.  
For applied fields between 10 and 20 mT, the coherence decay as $\tau$ is increased falls off roughly as $\exp[(\tau/T_2)^{-3}]$, with $T_2$ times of about $20\tau=60\,\mu$s.  Between 30 and 80~mT, the coherence persists out to progressively higher $\tau$ with applied field, and the decay curves acquire irregular bumps.  Although irregular, each bump is highly reproducible, at least over the many-minute timescales over which these measurements are taken.  
Demonstrating this are the two overlapping datasets at 40 mT, labeled \#1 and \#2, with highly reproduced features.  
Between 80 and 150 mT, the coherence decay regularizes again, albeit with a more exponential decay shape, and with greatly enhanced $T_2$ times of almost 1 ms.

The first hint that $^{73}$Ge underlies these phenomena is the observation that $T_2$ times from both CP10 and HE experiments appear to have a strong dependence on the width of the Si quantum well, which primarily impacts the amount of hyperfine coupling to $^{73}$Ge nuclei in the SiGe barriers. In Figure~\ref{fig:T2vWidthExp}, we summarize the $T_2$ times extracted from HE experiments conducted on several devices with varying growths of the Si well (the 5 nm data also appeared in the Supplement of~\cite{Eng2015}).  At low fields, the $T_2$ times increase monotonically with the width of the Si well from between 3 nm to 10 nm.  
The one exception to this is the low field $T_2$ for an 8~nm well device with isotopically natural silicon, whose growth conditions likely resulted in a more diffuse Si/SiGe interface than all the others, putting its characteristics similar to a device with a 3~nm well.
As the applied magnetic field increases into the 10s of mT, the $T_2$ times increase abruptly to approach 1~ms, with narrower well devices requiring larger applied fields before increasing. 

Also shown in Fig.~~\ref{fig:T2vWidthExp} are curves for Monte-Carlo simulations capturing the effects of quadrupole split $^{73}$Ge nuclei, which we discuss in more detail in Sec.~\ref{sec:t2}.  
Qualitatively, they behave similarly to the data at low magnetic field, showing a B-field-independent $T_2$ value depending on the well-width, and then diverging at some critical field.  
Lacking from these simulations but evident in the data is a high-field ``saturation" $T_2$ value.  
The experimentally observed $T_2$ saturation with field is likely limited by $^{29}$Si dipole-dipole dynamics, which are lacking in the simulations.  
We note that samples of widely varying well widths but all with the same Si isotopic enrichment have similar $T_2$ times at high field, and this $T_2$ increases gradually with field.  
Key evidence for this saturation being the $^{29}$Si limit is the two 8~nm well devices 
with natural abundance \isotope{Si}{29} in their wells, but differing Si/SiGe interfaces: both have much shorter high-field $T_2$ times of 100 $\mu$s.

\begin{figure}
	\includegraphics[width=\columnwidth]{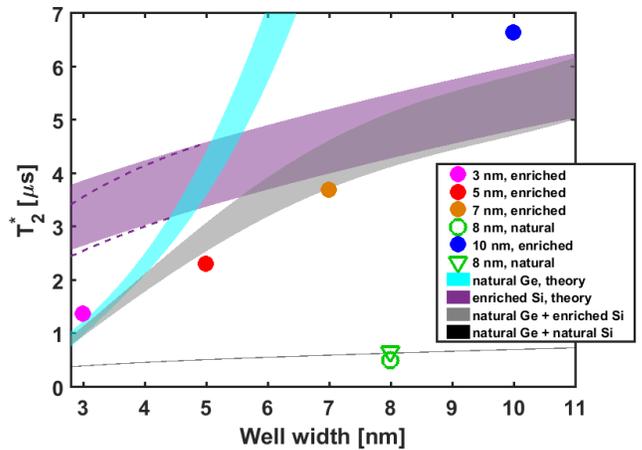}
	\caption{$T_2^*$ measured at a few mT for the same devices as in Figure~\ref{fig:T2vWidthExp}.  The two devices with natural isotopic Si content have $T_2^*$ values of 0.51 $\mu$s and 0.66 $\mu$s.  The colored bands show the $\mu\pm 2\sigma$ range, where $\mu$ is the mean value of $T_2^*$ defined for each nuclear species and $\sigma$ is the standard deviation under random isotopic placement, calculated from Eq.~(\ref{eq:T2stardef}).  The purple band shows the anticipated contribution of $^{29}$Si, while the cyan band shows the contribution of $^{73}$Ge, which dominates the amount of variation with well-width.  The gray bands are the total anticipated $T_2^*$.  The purple dashed lines show the small reduction of $T_2^*$ due to isotopically natural Si in the barriers; this reduction is negligible in comparison to the wells for wide well widths and in comparison to the Ge contribution for narrow well widths.}
	\label{fig:T2*_Width}
\end{figure}

The low-field $T_2$'s vary over nearly two orders of magnitude with well-width.  This contrasts sharply with the much weaker variation in $T_2^*$ with well width, as measured by FID: Fig.~\ref{fig:T2*_Width} shows the $T_2^*$ from the same devices as Fig.~\ref{fig:T2vWidthExp} at a few mT.  Unlike $T_2$, we see relatively minor variation of $T_2^*$ with applied magnetic field \cite{Eng2015}.   The variation we do see is discussed in Appendix~\ref{app:static_gradients}.  The $T_2^*$ values (Fig.~\ref{fig:T2*_Width}) correspond closely to predictions based on the anticipated overlap with nuclear spins of both species.  These predicted values of $T_2^*$ are dominated by $^{29}$Si at both natural abundance and at 800~ppm for wide wells, while $^{73}$Ge's contribution increases with narrower wells.  As the figure shows, the isotopic content of the silicon in the barrier plays a negligible role in comparison to either the natural \Ge\ spins in the barrier or the silicon in the well for all samples.  The ratio of magnitudes for the $T_2$'s as measured by HE at high field match the ratio of $T_2^*$'s for each device for wide wells, pointing again to $T_2$ at high field being dominated by the wavefunction overlap of $^{29}$Si.   

\subsection{Spectroscopic analysis}

\begin{figure*}
	\includegraphics[width=\textwidth]{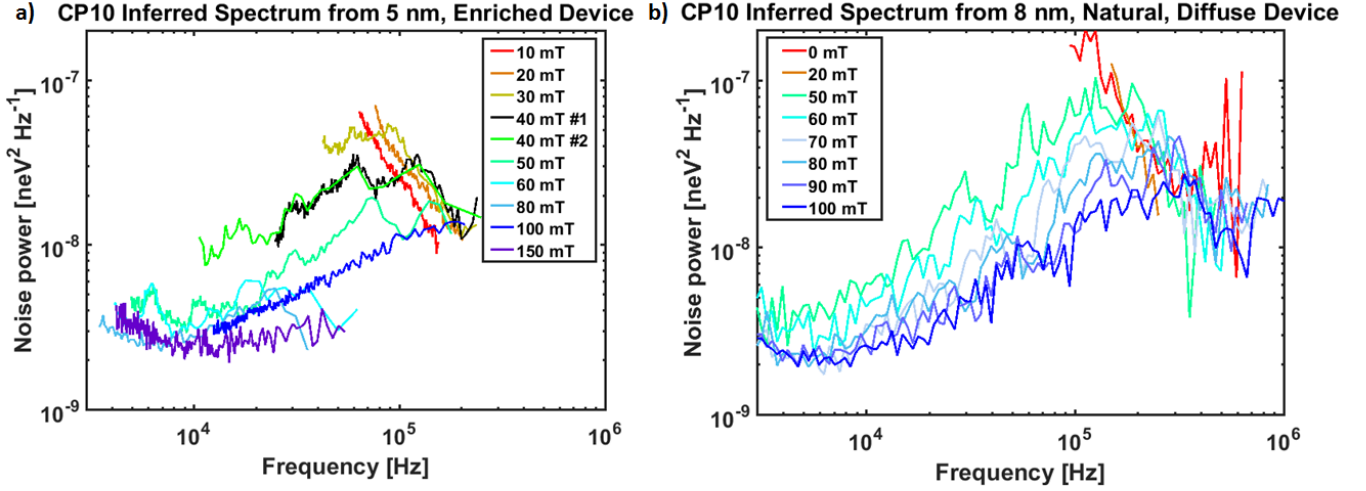}
	\caption{a) Inferred magnetic gradient noise from the CP10 data in Figure~\ref{fig:CP10vB} using Eq.~(\ref{inversion_definition}).  b) Same as (a), but from a device with an 8~nm well with natural \Si\ content and a more diffuse Si/SiGe barrier.}
	\label{fig:CP10_Spectra}
\end{figure*}

\begin{figure*}
	\includegraphics[width=\textwidth]{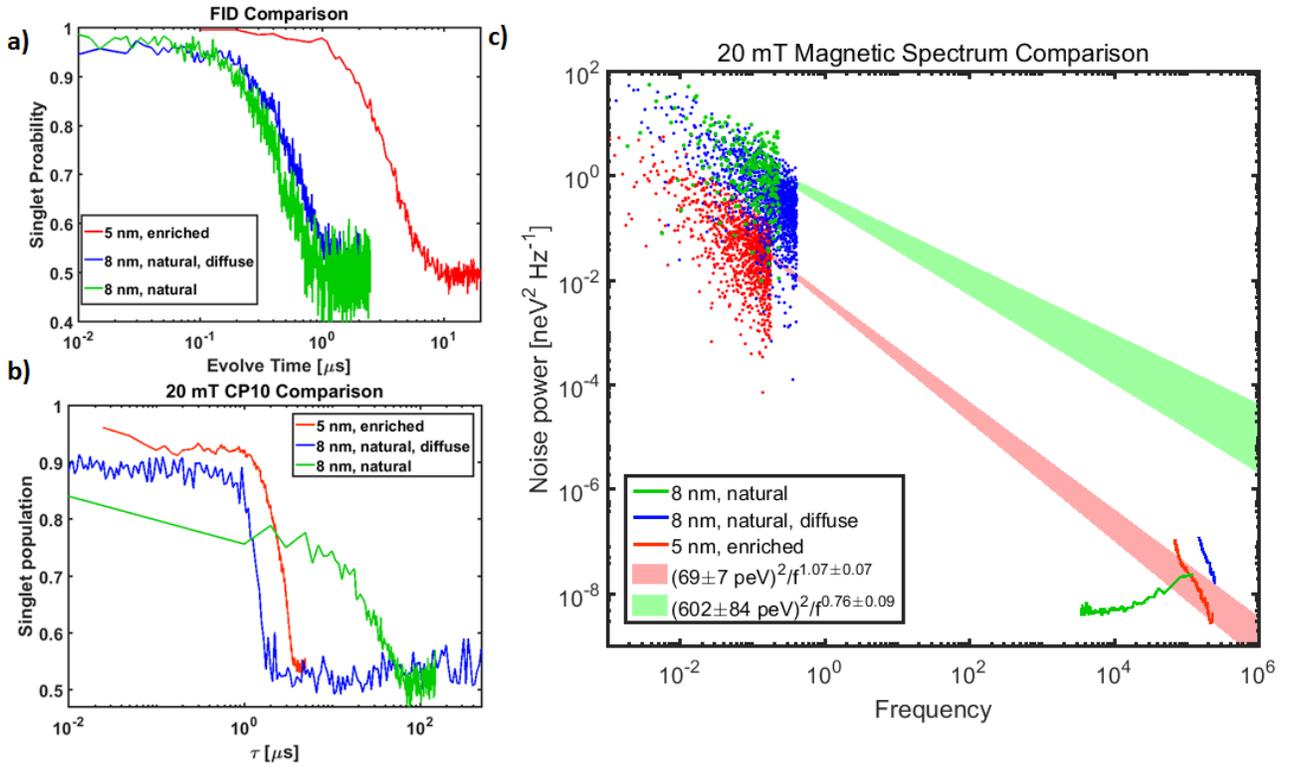}
	\caption{%
		a) FID experiments of the same devices at a few mT, showing that the \Si\ isotopic content has the largest contribution to $T_2^*$.
		b) CP10 decay experiments from three devices of differing material epitaxy, with 20~mT magnetic field, showing that the Si/SiGe interface has the largest contribution to CP10 decay.
		c) Comparison of spectral estimates from FID and CP10 experiments in 20 mT fields, as well as extrapolation of power law fits to the 5 nm and 8 nm CVD low frequency data.  The narrow, isotopically enhanced quantum well sample (red) has a clear connection between low and high frequency noise, as both have substantial \Ge\ contribution; the isotopically natural sample (green) has more low-frequency noise due to increased \Si\ content but comparable high-frequency noise due to comparable \Ge\ content.} 
	\label{fig:PSD_Comparison}
\end{figure*}

Both $T_2^*$ and $T_2$ indicate a strong role for $^{73}$Ge in relaxation at low field; experimental signatures of the quadrupole splittings of $^{73}$Ge are more evident by converting to the frequency domain using Eq.~(\ref{inversion_definition}), hence estimating the noise spectral density of magnetic gradient fluctuations from the CP10 data.  Using the time-domain data in Figure~\ref{fig:CP10vB}, Fig.~\ref{fig:CP10_Spectra}a shows the estimated magnetic noise spectra in this 5 nm enriched device~\footnote{ Figs.~\ref{fig:CP10vB},~\ref{fig:CP10_Spectra}, and~\ref{fig:PSD_Comparison} feature a different 5-nm-well sample than the one studied in Ref.~\onlinecite{Eng2015} and plotted in Figs.~\ref{fig:T2vWidthExp} and~\ref{fig:T2*_Width}}.
Spectral estimation only achieves a high signal-to-noise ratio for bandwidths over which the time domain signal has significant variation. Since the time domain data varies on the time scale $\tau$, the spectral estimation bandwidth is narrow and changes as $T_2$ changes.

We find that as the field increases, the overall noise magnitude decreases, while irregular peaks emerge, which flatten and increase in frequency.  At sufficiently high field (above 100 mT, where HE saturates to the $^{29}$Si limit) the spectra are much more flat.  Especially visible at fields 30 to 50 mT, a pair of spectral peaks appear to occur at the characteristic Larmor frequency of $^{73}$Ge, 1.5~MHz/T, and its first harmonic.  This intriguing feature signals the influence of quadrupole splitting, as we elaborate in the next section.

The estimated spectra in Fig.~\ref{fig:CP10_Spectra}a may be contrasted with that from CP10 data from devices with different material epitaxy, such as the
8-nm-well natural-Si device with diffuse sidewalls, Fig.~\ref{fig:CP10_Spectra}b.  
Here, we observe similar qualitative evolution, but with different magnitudes: the overall noise magnitude is higher and the noise peaks are less suppressed at the same applied fields.

The difference between the CP10-deduced spectra from different devices may be corroborated with the low-frequency noise estimates deduced from FID experiments.  
To illustrate, Figs.~\ref{fig:PSD_Comparison}a and b show time-domain comparisons of FID and CP10 experiments, respectively, for three samples.  
The $T_2^*$ of the device with a 5 nm well with enriched-\isotope{Si}{28} is longest at 2.9 $\mu$s, while the two 8 nm natural devices with sharper and diffuse interfaces have very similar $T_2^*$ values of 0.51 $\mu$s
and 0.66 $\mu$s, respectively.  
The underlying low-frequency ($<1$~Hz) noise spectra leading to these FID curves may be extracted using the method discussed in Ref.~\cite{Eng2015} and Appendix~\ref{app:static_gradients}, employing the static gradients available at 10-100~mT.  
These are plotted toward the left of Fig.~\ref{fig:PSD_Comparison}c.  
They have approximately $1/f$ power laws, as observed in all samples, and have total magnitudes consistent with the observed decays shown in Fig.~\ref{fig:PSD_Comparison}a.  
We may compare these $1/f$ spectra to the higher-frequency estimated spectra, shown in the lower right of panel c, extracted from the CP10 decay shown in panel b.  
While an extrapolation of the low frequency fit for the enriched 5 nm device matches the inverted high-frequency spectral band quite well, those of the 8~nm devices are much further off, again suggesting different mechanisms determine gradient noise in these two regimes.  
This may be understood by noting that $^{73}$Ge plays a substantial role in low-frequency noise (and thus $T_2^*$) for an enriched 5-nm well, but a negligible role in a natural 8-nm well; however, it plays a dominant role in the low-field $T_2$ spectrum in all three cases.  
This difference also points to why the two 8~nm devices may behave so differently: they have nearly the same $^{29}$Si character, which effects their low-frequency noise, but highly different $^{73}$Ge overlap, affecting their high frequency noise.

\section{Theoretical Model for Quadrupolar Coupling Determining $T_2$}
\label{sec:t2}

The experiments we have described already point strongly to $^{73}$Ge as a source of magnetic gradients limiting $T_2$ and $T_2^*$ in narrow wells, but they also show that the often uncertain details of the Si/SiGe interface across samples has significant effect on the measured values of decoherence times in FID, HE, and CP10 experiments.  
This may be understood from the form of the Fermi contact hyperfine interaction between the electron spin and the \Si\ and \Ge\ spins within the electron wavefunction in each dot.  
The contact hyperfine Hamiltonian for two well-separated quantum dots in the effective mass approach is
\be
\label{Hhf}
\ts{H}{hf} = \sum_{j=1}^{2} \sum_{k} \hbar A_{jk}\vec{S}_j\cdot\vec{I}_k,
\ee
where
\be
A_{jk} = \frac{2\mu_0}{3} g_0 \ts\mu{B}  \gamma_k \eta_k |\psi_j(\vec{r}_k)|^2.
\ee
The sum over index $j$ is over the pair of quantum dots, the sum over index $k$ is over all nuclear spins in the crystal, $\psi_j(\vec{r}_k)$ is the effective-mass envelope wavefunction for the electron in dot $j$ at the location of nucleus $k$, $\vec{S}_j$ is the spin of electron $j$ with vacuum electron gyromagnetic ratio $g_0\mu_B/\hbar$, and $\vec{I}_k$ is the spin of nucleus $k$ with gyromagnetic ratio $\gamma_k$.  
Note that $A_k$ is in units of rad/s here.
The overlap of the periodic Bloch wavefunction on each nuclear site, the ``bunching factors," are taken as unitless constants $\eta_k$.  
The total spin for each nuclear species are $I_k=1/2$ for \Si\ nuclei and $I_k=9/2$ for \Ge\ nuclei.

The values of $A_k$ may be approximately calibrated by assuming that FID measurements measure the high-field ergodic value of $T_2^*$, in which nuclear spins give a fair sample of their full distribution of configurations and the applied magnetic field causes electron Zeeman splittings significantly higher than hyperfine values.  
By increasing the averaging time and magnetic field in our experiments, we assure that our measured $T_2^*$ values are close to this limit.  
Under these assumptions, $T_2^*$ is estimated simply as 
\be
\left(
\frac{1}{T_2^*}
\right)^2 = \frac{1}{2}\sum_j\sum_k \frac{I_k(I_k+1)}{3}A_{jk}^2.
\label{eq:T2stardef}
\ee
To numerically estimate $T_2^*$ and use it to develop a barrier model, we require numeric values of the bunching factors $\eta_k$.  For Si, we use an $\eta$ value of 178, as this is well corroborated by calculation and magnetic resonance experiments~\cite{assali_hyperfine_2011}.  For Ge, the correct value to use is less clear~\cite{witzel_nuclear_2012}; our aggregated data across multiple devices with multiple quantum wells are consistent with a value of 570 which we assume throughout.  

With these values assumed we now note that the character of the Si/SiGe interface impacts the results in two ways.  
First, a diffuse interface allows the envelope wavefunction $\psi_j(\vec{r}_k)$ to broaden into the barriers more than it would for a hard-wall interface.  
Second, a diffuse interface places more straggling $^{73}$Ge in the quantum well region.  
Reference~\onlinecite{Dyck2017} show some example microscopic characterizations for some Si/SiGe wafers, and indicate clearly that interface diffusion is a significant consideration.  
Here, we find that our $T_2^*$ and $T_2$ data taken in aggregrate are decently supported by a model interface in which we simply convolve a perfect square interface with a Gaussian smearing function with standard deviation of 1 silicon mono-atomic-layer, 0.136~nm.  
Confounding our ability to estimate this interface from existing data is the uncertainty in the vertical electric field and the in-plane wavefunction diameter for each device.  We assume zero vertical electric field and a 30 nm $1/e$ diameter of a Gaussian $|\psi_j(\vec{r})|^2$, as these are consistent with electrostatic simulations, but these numbers are undoubtedly dependent on disorder and gate tuning variances.  
The resulting model for the barrier and wavefunction is illustrated in Fig.~\ref{fig:Dev}c for a 5~nm well.  
We emphasize that our estimated bunching factors, barrier model, wavefunction diameter, and vertical electric field are all rough estimates, enabling reasonably close postdictions for $T_2^*$ and $T_2$; in reality, the different samples under study may have different barrier and wavefunction parameters, so the assumption that all samples follow the same profile necessarily limits the accuracy of our models.

With the hyperfine hamiltonian so determined, we are now prepared to develop the theoretical model to explain not only the observed field and well-width dependences, but also the characteristic decay or spectral shapes of HE and CP10 experiments, arising from the electric quadrupole effects from the $^{73}$Ge nuclei.

The quadrupole moment of \Ge\ makes it quite different from \Si\, which is spin-1/2 and therefore only has a magnetic dipole moment.  
The result of the nuclear quadrupole moment of \Ge\ is to split the spin-states by an amount proportional to electric field gradients across the nucleus.  
The quadrupolar hamiltonian, Eq. VII.II.A.23 from Ref.~\onlinecite{Abragam1961}, is
\begin{multline}
	\label{HQ}
	H_Q=\sum_j\hbar\xi_j\biggl\{\frac{3\cos^2\theta_j-1}{2}[3{I^z_j}^2-I(I+1)]
	\\
	+3\sin\theta_j\cos\theta_j\{I^z_j,I^x_j\}
	+\frac{3\sin^2\theta_j}{2}[{I^x_j}^2-{I^y_j}^2]\biggr\}.
\end{multline}
Here, $\xi_j = e^2Q/[4I(2I-1)]\partial^2V/\partial z_j^2/\hbar,$ where $\theta_j$ is the angle between the $z$-axis (i.e. the applied magnetic field), and the principle axis of the electric field gradient tensor $\partial^2V/\partial x_j\partial x_k$ at the location of the nucleus.  
The $z$ direction is explicitly that of the magnetic field (not the crystal growth direction; in all experiments we have reported the magnetic field is in the plane of the quantum well).  
The electric quadrupole moment of the $^{73}$Ge nucleus is defined in terms of $eQ=\langle \sum_j e_j(3z_j^2-|\vec{r}_j|^2)\rangle$, where the sum is over nucleons and the expectation is over the maximal spin state $m_z=I$.  While the quadrupole moment $Q$ for \Ge\ is known (about $-200$~millibarn), the electric field gradients contributing to quadrupolar splittings are far more uncertain.

We know of no direct measure or calculation of \Ge\ quadrupole splittings in SiGe alloys.  
We note that the symmetry of a bulk silicon crystal means that an isolated \Ge\ in a perfect silicon host would see no quadrupole splitting.   
However, in the SiGe barrier or at its interface, the relevant nuclei see an electric environment with substantially reduced symmetry, due in majority part to alloy effects (i.e. the neighbors of a given \Ge\ are random combinations of Ge and Si) and to a much lesser degree due to strain.   
For alloy effects, Ref.~\onlinecite{wright_mossbauer_2016} reports calculations of quadrupole splittings for a substitutional iron atom in a Si crystal, in the presence of a perturbation due to the Ge atom nearby; these calculations inform about plausible magnitudes of quadrupole splittings, and tabulate the effect of first, second, and third nearest neighbor Ge atoms.  
In the barrier, with 30\% Ge, most (76\%) of lattice sites have at least one 1st nearest neighbor Ge.  
While the effect of a 2nd nearest neighbor Ge is reported to be about four times weaker than a first nearest neighbor, there are 3 times more of them, resulting in a significant possible range of quadrupole splittings in the alloy.  
We estimate that these alloy effects, using Ref.~\onlinecite{wright_mossbauer_2016} as proxy, overwhelm the effects of macroscopic strain present in the device due to the lattice mismatch of the Si/SiGe heterostack, as well as internal stresses due to thermal expansion mismatches of the gatestack during cool-down; using the similar strain-to-quadrupole calibrations available from studies of $^{75}$As~\cite{franke_interaction_2015} and $^{123}$Sb~\cite{asaad_coherent_2020}.  
Existing data on quadrupole splittings in other semiconductors may also inform estimates for the magnitude of the $\xi_j$ constants. 
Reference~\onlinecite{Verkhovskii2003} reports quadrupole splittings from nuclear magnetic resonance (NMR) studies of single-crystal Ge with varying isotopic contents, attributing electric field gradients to isotopic disorder.  
In this case, quadrupole splitting parameters $\xi_j$ are on the order of 100~Hz.  
A closer estimate results from noting that AlGaAs/GaAs quantum wells have similar alloy fluctuations to Si/SiGe quantum wells and the \isotope{Ga}{69},\isotope{Ga}{71}, and \isotope{As}{75} nuclei have similar quadrupole moments $Q$ as \Ge.   
Since NMR spectra are available from these quantum wells via optical detection~\cite{guerrier_calibration_1997}, quadrupole splittings in the 1 to 30~kHz are observable and suggest a comparable range for SiGe.  

A characteristic feature of $H_Q$ is the existence of matrix elements allowing $\Delta m_z = 0,\pm 1,$ and $\pm 2$.  
These transitions correspond to shifts of the nuclear Zeeman energy, described by hamiltonian
\be
\label{HZ}
H_Z = \sum_k \hbar\gamma_k B_0^z I^z_k,
\ee
where the \Ge\ nuclei each have $\gamma_k=\gamma_{73} = (2\pi)\times 1.49$~MHz/T.  In the interaction picture for this Hamiltonian, the quadrupole Hamiltonian $\tilde{H}_Q(t) = \exp(iH_Zt/\hbar)H_Q\exp(-iH_Zt/\hbar)$  features terms at zero-frequency, the nuclear Larmor frequency $\gamma_{73} B_0^z$, and twice the nuclear Larmor frequency $2\gamma_{73} B_0^z$; these terms also provide matrix elements to vary the coupling to electrons $H_{\text{hf}}$.  Roughly speaking, then, we anticipate this ``bath" to provide ``noise" at these characteristic frequencies.  This qualitative observation is somewhat evident in Fig.~\ref{fig:CP10_Spectra}, where two peaks are visible at 30-50~mT range coincident with $\gamma_{73} B_0^z$ and $2\gamma_{73} B_0^z$.  

To make a more quantitative assessment, we simulate the dynamics of randomly precessing and nutating quadrupolar \Ge\ and track their hyperfine shifts.  
The noise source in question is highly non-Markovian, since in the regime that \Ge\ hyperfine coupling constants are comparable to the \Ge\ Larmor frequency, the evolution of quadrupolar nuclei will be highly influenced by the electron spin-flips which occur in each dot due to exchange pulsing.  
The act of swapping electron-spins, which amounts to flipping the hyperfine field seen by each \Ge\ nucleus, may be regarded as a periodic variation of the effective Larmor frequency, and hence modifies the noise dynamics as translated through the hyperfine interaction.  
For this reason, in addition to the complication that all terms of the Hamiltonian are of comparable order in those regime where the most prominent spectral features occur, perturbative expansions are challenging for analysis.  
Fortunately for calculation of quadrupolar effects, much of the structure of the CP$n$ observations may be understood considering only non-interacting \Ge\ nuclei.
The resulting motion may therefore be simulated by exponentiating the 10-dimensional spin Hamiltonian for each \Ge\ and for each electron spin orientation, composing unitary operators by alternating the sign of the electron spin, and summing the contributions of each nucleus on the electrons.  
The detailed equations describing the simulations of $^{73}$Ge precession amongst spin swaps is presented in Appendix~\ref{sec:CPnsim}.  

\begin{figure*}
	\includegraphics[width=6in]{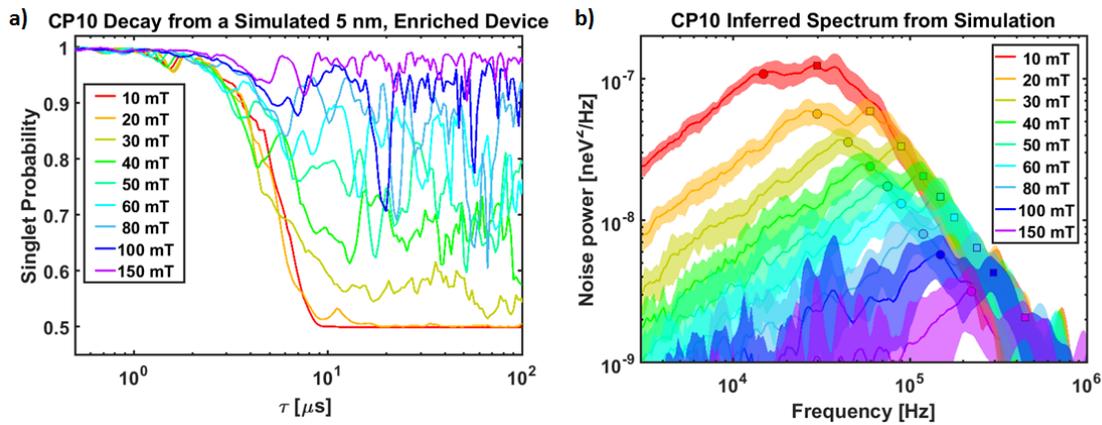}
	\caption{a) Simulated decay of a CP10 experiment vs. magnetic field under the effects of \Ge\ quadrupolar dynamics only, for a single simulated device with a 5-nm-wide, isotopically enriched Si well.  Nuclear dipole-dipole dynamics are not included in this simulation. b) Inferred magnetic noise spectrum, via Eq.~(\ref{inversion_definition}), from an ensemble of 20 time-domain simulations such as that in (a), each using a different random configuration of \Ge\ nuclei.  The lines show the average spectrum across the ensemble, and the bands the standard deviation.  The circles on each trace show the location of the \Ge\ Larmor peak and the squares its first harmonic.  }
	\label{fig:CP10vBsim}
\end{figure*}

An example simulation of a CP10 experiment for a 5-nm-well device, including only $^{73}$Ge spins which are randomly placed in a simulated crystal, is shown in Figure~\ref{fig:CP10vBsim}(a).   
While the detailed structure of the decay will depend on the specific placement of \Ge\ nuclei in the barrier of the device, we find this non-interacting simulation well-simulates the initial structure of the CP$n$ decay, as evident from a qualitative comparison to Fig.~\ref{fig:CP10vB}.  
For this simulation, we find that drawing the hyperfine splitting constants $\xi_j$ from a Lorentzian distribution with zero mean and root-mean-square width of 10~krad/sec gives decay characteristics comparable to experiment; varying this width by up to an order of magnitude around this estimate gives insignificant variation relative to the variation resulting from random isotopic placement.  
Figure~\ref{fig:CP10vBsim}(a) fails to show the smooth Gaussian decay at high field or at high values of $\tau$, since it results from a non-interacting, ultimately fully coherent model.  
To capture this additional decay, we would have to include the $1/f$ noise believed to result from nuclear dipole-dipole interactions.  
As this results from an extended many-body system, comparable simulations would be more complex, requiring methods such as coupled-cluster expansion techniques ~\cite{Witzel2005,witzel_converting_2014} not employed here.  

The resemblence of this simple quadrupolar model to experiment is perhaps more evident using the noise spectrum inferred from data and now from simulation, shown in Fig.~\ref{fig:CP10vBsim}(b).  
In this case, we simulate CP10 experiments for 20 independent simulated 5-nm-well samples (with \Ge\ isotopic placement in the crystal randomized for each sample) and plot the average and standard deviation of the inverted noise spectrum, for comparison to Fig.~\ref{fig:CP10_Spectra}.  
Here, we see that our simple model of non-interacting, quadrupole-split $^{73}$Ge nuclei result in broad noise peaks at $\gamma_{73} B_0^z$ and $2\gamma_{73} B_0^z$ with noise power at comparable values to that in Fig.~\ref{fig:CP10_Spectra}(a), with similar reduction in power with applied field $B_0^z$.  
Again, the low-frequency noise content varies from experiment due to the lack of nuclear dipole-dipole interactions; Fig.~\ref{fig:PSD_Comparison} suggests a model in which this additional noise source can be measured (or estimated) at low-frequency and simply added to the \Ge\ noise peak.

This same simulation, now using only a single exchange-swap, may be used to understand the magnetic field variation of Hahn-echo experiments as well.  
The thin lines in Fig.~\ref{fig:T2vWidthExp} are the result of Hahn echo simulations from 10 isotopically randomized crystals at each of the 5 well widths considered, again determining $T_2$ as that particular value of $2\tau$ where the singlet probability falls to $1/e$.  
Comparable to the data, the simulations indicate that $T_2$ due to non-interacting quadrupolar nuclei is roughly constant at low field and then diverges at some critical field $B_0$.  
The location of this ``critical" field corresponds to where the \Ge\ Larmor frequency begins to significantly exceed both the broad hyperfine coupling parameters $A_{k}$ and the quadrupole splittings $\xi_j$, which are of comparable magnitude 
\footnote{In Ref.~\onlinecite{Eng2015}, we had observed this critical field and indicated its proximity to where the electron Zeeman frequency matches the electron temperature, but indicated the material origin was unknown; we now consider that energetic proximity a coincidence, and point to \Ge\ instead to explain the $T_2$ vs. $B_0^z$ in that prior work.}.
At much higher fields, non-interacting \Ge\ spin projections freeze, and remaining noise properties would have to result from spin flip-flops, especially \Si\ dipole-dipole dynamics as we have previously discussed, and are excluded from the simulations shown in Fig.~\ref{fig:T2vWidthExp}.  
As with the CP10 simulations, the quantitative results of $T_2$ vary weakly with the unknown quadrupole coupling parameters $\xi_k$; the more critical variation is the microscopic character of the Si/SiGe barrier profile.  
An interesting open question is whether more detailed quantitative fitting of data such as we have shown, or the results of more complex noise spectroscopy sequences beyond what we have shown here, may allow more precise determination of barrier profiles.

\section{Discussion}

The magnetic noise effects reported here are more apparent relative to prior quantum dot studies in part due to the use of isotopically enhanced Si (i.e. 800 ppm \Si).
In nuclear-rich materials such as GaAs, spin-orbit and quadrupolar effects are also incidentally larger and so may be observed using Landau-Zener transitions~\cite{dickel_characterization_2015,pal_electron_2017} and dynamical decoupling~\cite{botzem_quadrupolar_2016,malinowski_notch_2017,nakajima_coherence_2020}. 

We emphasize, though, that a key reason the noise sources are evident in the devices under discussion in the present study is that these devices are designed to \emph{minimize} magnetic gradients, either from nuclear spins or micromagnets, in order to achieve exchange-only control, whose construction relies on magnetic field homogeneity~\cite{DiVincenzo2000,Eng2015,Andrews2019}.  Many Si/SiGe devices presently under study include magnetic field gradients, and in these, quadrupolar effects may be difficult to observe due to the interference of charge-noise~\cite{Yoneda2018,struck_low-frequency_2020}. The singlet-triplet measurement scheme further allows us to study a wide range of low magnetic fields, not available in single-spin readout schemes~\cite{fogarty_integrated_2018,zhao_single-spin_2019}.  

Exchange-only operation is particularly sensitive to magnetic noise due to the possibility of leakage, even while pulsing~\cite{ladd_hyperfine-induced_2012,Reed2016,Andrews2019}.  
The application of correction sequences may still allow very high fidelity operation, only insofar as the operation time well exceeds high-frequency cut-offs of the magnetic noise~\cite{wang_noise-compensating_2014,throckmorton_fast_2017}.  
For this, the observation of higher frequency \Ge\ quadrupolar noise peaks reported here is most pertinent, but optimistically we may be confident that dipolar noise cannot extend to arbitrarily high frequencies and \Ge\ quadrupolar peaks may easily be circumvented.  
We find that the effects of \Ge\ quadrupolar dynamics diminish significantly with field, yielding the long, \Si-limited $T_2$ times at easily operational magnetic fields.

Recently, quadrupolar spins have shown surprising resonant response to gate voltage in donor devices~\cite{asaad_coherent_2020}.  
We neither see nor expect evidence of such effects in the present results, but under different designs, we speculate that the electric quadrupole effects of \Ge\ in SiGe might serve as an interesting lever rather than a decoherence source in future silicon spin qubit devices.

In summary, we have argued that low-field $T_2$ in exchange-only qubits in the absence of magnetic field gradients results from electrical quadrupole effects from the \Ge\ nuclei, as corroborated by the correct well-width, $B$-field, and decay-shape dependence relative to theoretical models.  
The strong correspondence is reassuring for the future of exchange-only qubits in this material, as it suggests this noise can be circumvented by reasonable applied magnetic fields, by noise compensation techniques performed at moderate speeds, or by depletion of the \Ge\ spin isotope in the SiGe material.  Nonetheless, additional research remains in the modeling and detection of the magnetic noise features studied here to solidify predictions for the ultimate performance of SiGe qubits.   

\acknowledgements{
We acknowledge valuable experimental contributions from %
Chris~Bohn, %
Devin~Underwood, %
Laura~De~Lorenzo, %
Ed~Chen, %
Cathie~Erickson, %
Mitch~Jones, %
Ari~Weinstein, %
Matt~Reed, %
Reed~Andrews, %
Matthew~Rakher, %
Matt~Borselli, %
and Andy~Hunter; %
valuable theoretical discussions with 
Matthew~Grace, Wayne~Witzel, Seth~Merkel, and Chris Schnaible; %
and assistance with figures from John~B.~Carpenter.}

\appendix

\section{Filter Function Formalism}
\label{app:FF}

In all experiments, a singlet is prepared in the well-separated double-dot system, and the measurement samples the probability of ending in this same state.
In between, for each ensemble instance of an experiment, each spin is presumed to evolve independently in each dot, resulting in the unitary
where $\vec{S}_j$ is the vector spin operator for electron $j$, and $\vec{b}_j(t)$ is a time-dependent, noisy vector.
This $\vec{b}_j(t)$ tracks the fluctuating magnetic field, for which we assume there may be a static component, notated as $B_0$, and vector fluctuating components $\delta\vec{B}(t)$.
The filter-function formalism is ultimately a linear expansion with respect to this fluctuation, and hence employs a first order Magnus expansion for $\vec{b}_j$:
\be
b_j^\alpha (t) \approx \frac{g\ts\mu{B}}{\hbar}\int_0^t ds \ h_{jk}(s) R(t)^{\alpha\beta} B^\beta_k(s),
\ee
where repeated indices are summed, superscripts correspond to the three spatial vector components, and the subscripts refer to which dot the electron occupies during the experiment.
Here $g$ is electron $g$-factor, $\ts\mu{B}$ the Bohr magneton, and $\vec{B}_k(t)$ is the time-dependent magnetic field in dot $k$.
The matrix $\vec{R}(t)$ rotates vectors about the magnetic field axis at the angular Larmor frequency $\omega_0 = g\mu_B B_0/\hbar$.
The time-dependent ``switching matrix" $h_{jk}(t)$ tracks how spins are swapped between dots during the evolution.
For the three experiments in this paper, the switching matrices are
\begin{align}
	\text{FID}: h(s) &= \begin{pmatrix} 1 & 0 \\ 0 & 1\end{pmatrix}
	\\
	\text{HE} : h(s) &= \begin{cases}
		\begin{pmatrix} 1 & 0 \\ 0 & 1 \end{pmatrix}, & 0 \le s < \tau \\
		\begin{pmatrix} 0 & 1 \\ 1 & 0 \end{pmatrix}, & \tau \le s < 2\tau
	\end{cases}
	\\
	\text{CP}n : h(s) &= \begin{cases}
		\begin{pmatrix} 1 & 0 \\ 0 & 1 \end{pmatrix}, & 4m\tau \le s < (4m+1)\tau \\
		\begin{pmatrix} 0 & 1 \\ 1 & 0 \end{pmatrix}, & (4m+1)\tau \le s < (4m+3)\tau\\
		\begin{pmatrix} 1 & 0 \\ 0 & 1 \end{pmatrix}, & (4m+3)\tau \le s < 4(m+1)\tau
	\end{cases},
\end{align}
where in the case of CP$n$ ($n$ even), $m$ is an integer ranging from 0 to $n/2-1$, capturing the repetition of this experiment.
Note that these matrices are necessarily centrosymmetric.
As a result of the unitary evolution for each experiment, the probability of measuring the singlet in a single experiment is
\begin{multline}
	|\!\bra{S}U(t)\ket{S}\!|^2 =
	\\
	\left(\cos\frac{|\vec{b}_1|}{2}\cos\frac{|\vec{b}_2|}{2}
	+\frac{\vec{b}_1\cdot\vec{b}_2}{|\vec{b}_1||\vec{b}_2|}
	\sin\frac{|\vec{b}_1|}{2}\sin\frac{|\vec{b}_2|}{2}\right)^2.
\end{multline}
Note that in the case that all magnetic fields are parallel, in which case $\vec{b}_j$ may be treated as a scalar $b_j$, this expression simplifies to
\be
|\!\bra{S}U(t)\ket{S}\!|^2 \rightarrow \frac{1}{2}+\frac{1}{2}\cos[b_1(t)-b_2(t)], \quad \vec{B}_1 || \vec{B}_2.
\ee

We now average these expressions over noisy magnetic fields, assuming the noise is stationary and Gaussian.
Again consistent with the first-order filter-function expansion, we limit our derivation to the first and second moments, allowing the estimated evolution of \refeq{PS}, which would be exact for the all-parallel-field case under strict Gaussian assumptions.
In the case of non-parallel fields with non-zero mean, the constant field $B_0$ is taken along the difference vector $\vec{B}_1-\vec{B}_2$.
To evaluate the second moment, we make the highly simplifying assumption that different vector components of noise within each dot are uncorrelated and identical, and that noise in each dot is uncorrelated and identical:
\be
\langle B_j^\alpha (t_1) B_k^\beta (t_2) \rangle = \delta^{\alpha\beta}\delta_{jk} \int_0^\infty df S_j(f) \cos[2\pi f(t_1-t_2)],
\ee
where $S_j(f)$ is the magnetic noise spectral density for dot $j$.
Under this simplification, we find \refeq{FFdef},
where the filter function $F(f,t)$ may be divided into a central lobe $F_0(f,t)$ and two side-lobes $F_\pm(f,t)=F_0(f\pm f_0,t)$; i.e. the side-lobes are separated from the central lobe by the Larmor frequency $\omega_0/(2\pi)=g\ts\mu{B}B_0/h$:
\be
F(f,t) = F_0(f,t) + F_0(f+ f_0,t)+F_0(f- f_0,t).
\ee
The central lobe is derived by dividing an experiment into $N$ intervals, such that $t=N\tau$, and writing
\be
F_0(f,t)=\sum_{p,q=1}^N C_{pq}\frac{\cos[2(p-q)\pi f\tau]\sin^2(\pi f\tau)}{(\pi f)^2},
\ee
where the matrix $C_{pq}$ is given by
\be
C_{pq} =\frac{1}{2}( h_{1k,p} h_{1k,q} - h_{1k,p} h_{2k,q} - h_{2k,p} h_{1k,q} + h_{2k,p} h_{2k,q}),
\ee
where $h_{jk,p} = h_{jk}(s)$ for $(p-1)\tau \le s < p\tau$. 
For the three experiments in this manuscript, the matrix $C_{pq}$ is very simple.
For FID, $N=1$ and $\ts{\vec{C}}{FID}=1$.
For HE, $N=2$ and $\ts{\vec{C}}{HE}$ is one for both diagonal components and $-1$ for both off-diagonal components.
For CP$n$, $N=2n$ and the matrix is full of $n/2\times n/2$ repeated $4\times 4$ blocks, each of which is $\ts{\vec{C}}{HE}\otimes \ts{\vec{C}}{HE}$.
We note that the present formalism may be easily extended to more dots and more complex sequences, in which case the $\vec{C}$ matrices become more complex.

For FID, that central band has the filter function
\be
\label{FIDFF}
F_0(f,t) = \frac{\sin^2(\pi f t)}{(\pi f)^2}
\ee
while HE and CP$n$ experiments have the central band described as Eqs.~(\ref{HEFF}) and and (\ref{CPnFF}), respectively.
A key observation from our approximations is that the filter function $F(f,t)$ contains not only the central lobe $F_0(f,t)$ above, which is independent of $B_0$, but also two equivalent sidelobes separated by the Larmor frequency $f_0$.
We can keep only the central band if either (1) all magnetic fields are parallel or (2) if the Larmor frequency is much higher than all constituent noise frequencies of the underlying magnetic field fluctuation.
For finite external field and low frequency magnetic noise, the FID decay including the Larmor sidelobes takes the form
\be
\label{eq:lowfieldfid}
P_S(t) \approx \frac{1}{2}+\frac{1}{2} \exp\left(-\frac{t^2 \omega_0^2+4(1-\cos \omega_0 t)}{{T_2^*}^2 \omega_0^2 } \right),
\ee
reducing to the standard expression in the $B_0\rightarrow\infty$ limit. 

\section{Static and Low-fluctuation-frequency Field Gradients}
\label{app:static_gradients}

\begin{figure}
	\includegraphics[width=0.9\columnwidth]{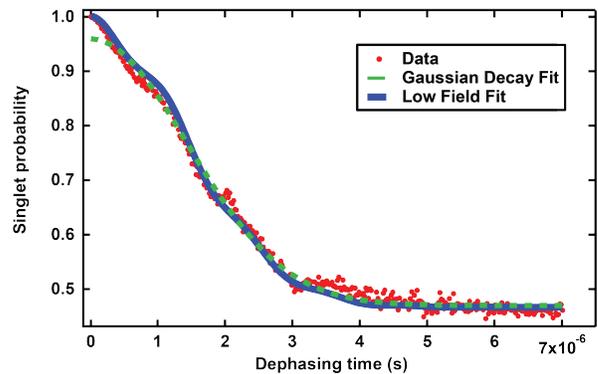}
	\caption{
		Free induction decay taken at very small external magnetic field.
		The dashed green line is a fit to a Gaussian decay curve, resulting in a $T_2^*$ = 2.04~$\mu$s.
		The solid blue line is a fit to Eq.~(\ref{eq:lowfieldfid}) and results in a $T_2^*$ = 1.98$\mu$s with a magnetic field of 29.8~$\mu$T. }
	\label{fig:FID}
\end{figure}

If no exchange echo pulses are applied, the simple decay experiment is referred to as Free Induction Decay (FID), in analogy to the common experiment in NMR.  
Since this experiment has a central-lobe filter function, Eq.~(\ref{FIDFF}), that is finite at $f=0$, it is sensitive to very slow changes in magnetic gradients.  
Those magnetic field gradients, which arise due to fluctuating nuclear spins in both isotopically natural and enhanced samples, are well described as approximately $1/f$ Gaussian noise, resulting predominantly in a Gaussian decay, as in Fig.~\ref{fig:PSD_Comparison}(b).  
However, we also observe structure in the FID decay in some magnetic field regimes, which we discuss in this appendix.

Some structure is observed at very low magnetic field, such as the Earth's magnetic field, due to the influence of transverse magnetic field fluctuations.  
In Fig.~\ref{fig:FID}, we show a well-averaged FID curve in a device with a 3-nm quantum well taken at Earth's magnetic field.  A fit of the oscillations superimposed on the Gaussian decay is highly consistent with the field-dependent filter function as derived in Appendix~\ref{app:FF}, fitting to a total applied magnetic field of $29.8\pm0.04$~$\mu$T, consistent with the Earth's magnetic field.  

\begin{figure*}
	\includegraphics[width=0.7\textwidth]{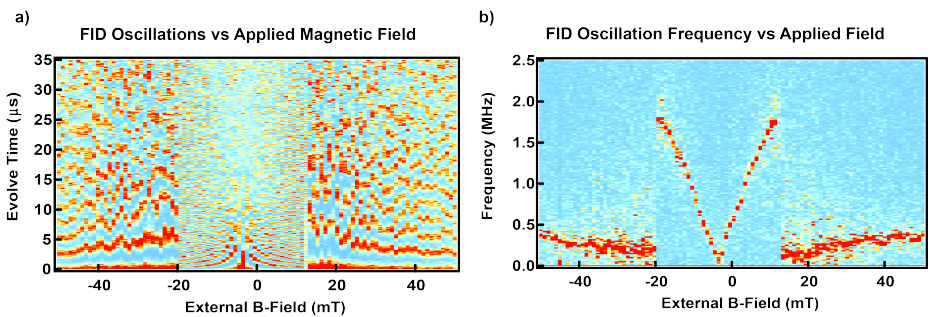}
	\caption{Free induction decay versus magnetic field.  (a) FID decay in the time domain, showing sinusoidal variation under a Gaussian envelope as in Eq.~(\ref{PS}).  (b) Fourier transform of (a), showing that the frequency of the singlet-tripet oscillations increase with applied field in two varying regimes.  Note that the applied field is taken from a calibration of the current in the power supply of the superconducting solenoid, which typically carries a few-mT offset from true zero field, evident in this data.}
	\label{fig:Meissner}
\end{figure*}

At high magnetic fields, static gradients appear, as evident from singlet-triplet oscillations indicated as the $\langle|\vec{b}_1(t)-\vec{b}_2(t)|\rangle$ term of Eq.~(\ref{PS}).  
An example such curve was presented in Ref.~\onlinecite{Eng2015}, Fig. 6A, and a series of such curves is shown in Fig.~\ref{fig:Meissner}(a).  
In both cases, we have observed that singlet-triplet oscillations have a frequency increasing linearly with applied magnetic field.  
The samples in the present study, however, see larger gradients at smaller fields. 
Figure \ref{fig:Meissner}(b) shows the Fourier transform of the singlet-triplet oscillations, whose peaks are comparable to Fig. 6B of Ref.~\onlinecite{Eng2015}.  
In the present study we clearly see two regimes; at a critical field of about 18~mT, we see the oscillations suddenly reduce in frequency.  
The higher frequency, low-field paramagnetic singlet-triplet oscillations were not observed in the sample of Ref.~\onlinecite{Eng2015}, which employed Ti/Au gates.  
The devices in the present study use aluminum gates, suggesting the superconductivity of aluminum at the 50~mK base temperature of these experiments is at play.  
Moreover, the critical field is found to be mildly hysteretic, characteristic of near-critical superconducting effects in aluminum thin-films.  
In a study on another series of samples, we have validated that for the method and thickness of aluminum deposition we employ, the critical field for superconductivity is highly consistent with the observed field at which the large singlet-triplet oscillations reduce.  
We therefore posit that the static magnetic field gradient at low-field occurs due to screening of the applied magnetic field from the Meissner effect in the aluminum gates.  
We have further performed electrostatic modeling of the field profile resulting from the screening expected from perfect Meissner gates in our gate geometry and found it consistent with the magnitude of the gradients observed.

At higher magnetic fields (e.g. $>$18 mT in Fig.~\ref{fig:Meissner}), and in previous devices with Ti/Au gates, the singlet-triplet oscillations are similar to those 
also observed in silicon MOS devices~\cite{Jock2018,tanttu_controlling_2019}.   These oscillations are now attributed to spin-orbit coupling at the barrier Si/SiO$_2$ or the Si/SiGe barrier, as evidenced by their characteristic variation with magnetic field orientation~\cite{ferdous_interface-induced_2018}.

Due to the broad filter function of the time-averaged FID experiment, it is not possible to reconstruct the noise spectrum which gives rise to the Gaussian decay using only FID decay curves.  Instead, we use the sinusoidal shape of the singlet-triplet oscillations due to Meissner screening or spin-orbit effects to perform peak-tracking.  By fitting a time-series of FID experiments to Eq.~\ref{PS}, we may track the location of the phase of the fixed-frequency oscillation and use this to extract the low-frequency character of gradient noise.  This technique was previously described in Ref.~\onlinecite{Eng2015} (See Fig.~S1 there),
but in that work, larger magnetic field magnitudes had to be used to induce oscillations, whereas the data presented in Fig.~\ref{fig:PSD_Comparison} was taken at an applied field of 20~mT.

As shown in Fig.~S1 in Ref.~\onlinecite{Eng2015} and in Fig.~\ref{fig:PSD_Comparison} of the present work, all silicon samples show a nearly $1/f$ spectrum at fields 10s of mT.  This contrasts the most basic models of spin diffusion which indicate a $1/f^2$ spectrum.  Some recent evidence for $1/f$-like behavior is also seen in the modeling of P:Si donor devices~\cite{madzik_controllable_2020}.  We posit that in both isotopically natural and 800~ppm \Si\ devices, the $1/f$ noise results from the large distribution of dipole-dipole timescales resulting from random \Si\ placement.  Subdiffusive behavior has similarly and recently also been reported in GaAs dipolar dynamics in a single quantum dot~\cite{nakajima_coherence_2020}.  Detailed time-domain simulations of $^{29}$Si dynamics in quantum dots~\cite{witzel_converting_2014} are needed to verify the underlying reasons for this particular power law.  The present study at least verifies that the origin of this $1/f$ noise is indeed nuclear in origin (in contrast to $1/f$ magnetic noise arising from charge noise in samples including micromagnets~\cite{Yoneda2018,struck_low-frequency_2020}), as its amplitude varies as expected with isotopic content and with quantum well width.

\section{Simulation of \Ge\ Quadrupole Effects under Exchange Echo Experiments}
\label{app:Ge}

\label{sec:CPnsim}

Here we derive the model used to simulate \Ge\ dynamics during multiple-exchange-echo experiments for a singlet-triplet qubit in a double quantum dot.  
For the singlet-triplet subspace of the two electrons in the double dot, we may neglect hamiltonian components proportional to the total spin-projection operator $S_1^z+S_2^z$ since at fields $>$ 1 mT we presume to stay in the null space of this operator.  
We are most concerned with the singlet-triplet qubit $\sigma_x$ operator $\sigma^x = S_1^z-S_2^z$, since it is magnetic field differences between the two dots that drive transitions between the singlet-triplet qubit states.  We may therefore write the hyperfine hamiltonian as
\be
\tsc{H}{hf} \approx  \frac{\sum_{k}A_{k}I_k^z-\sum_{k'}A_{k'}I_{k'}^z}{2}\sigma^x,
\label{SThf}
\ee
where the sums over $k$ and $k'$ are the sums over the nuclei in the two dots, and $\sigma^x$ drives singlet-triplet oscillations.  (We have hence altered our notation relative to Eq.~(\ref{Hhf}) to drop the $j$ (dot) subscript on $A_{jk}$, and instead track ``which dot" of $A_k$ by the location of nucleus $k$.)

Our basic model is that each nucleus, indexed by $k$ and $k'$ in Eq.~(\ref{SThf}), precesses in its individual hyperfine environment and interacts only with the common electrons of the quantum dot.  
Nuclear polarization must reconfigure due to some interaction (dipole-dipole, quadrupole relaxation, hyperfine-mediated interactions, etc.) but we treat these reconfigurations as quasistatic; i.e. slow in comparison to a single spin-echo measurement, albeit fast enough to observe on the averaging timescale of a time-ensemble of measurements.  
We also treat the nuclear spin bath as having effectively infinite spin temperature; $\gamma_{73}B_0^z$ is smaller than thermal energies at our applied fields, but more critically nuclear $T_1$ at mK would not allow thermalization to a finite temperature even at the longest averaging times in the experiments described.  
These assumptions, common in analyses of nuclear baths, enable us to simply trace over the nuclear degree of freedom when predicting a measurement result.

To analyze decoherence, we consider evolution in the two subspaces which are eigenstates of $\sigma^x$, i.e. $\ket{+}$ and $\ket{-}$.  
(These states in fact correspond to $\ket{\uparrow\downarrow}$ and $\ket{\downarrow\uparrow}$ for electron-spins in the double dot).  
Hence we may decompose the relevant terms of our total hamiltonian as
\be
H = \sum_k H^+_k\ketbra{+}{+}+H^-_k\ketbra{-}{-},
\ee
where
\be
H^\pm_k = H_Z+H_Q\pm \frac{s_k}{2}A_kI_k^z,
\ee
acts on only nucleus $k$.  
Here $s_k=\pm 1$ depending on whether nucleus $k$ is from dot 1 or dot 2.

Evolution for a time $\tau$ under one of these Hamiltonians is calculated via simple exponentiation, as
\be
U^\pm_k = e^{-iH^\pm_k\tau/\hbar}.
\ee
In an exchange-echo experiment, we begin with a singlet state $\ket{S}=(\ket{+}-\ket{-})/\sqrt{2}$.  
We evolve for time $\tau$ to state
\be
\ket{\psi(\tau)} = \frac{1}{\sqrt{2}}\biggl[\prod_kU^+_k\ket{+}-\prod_kU^-_k\ket{-}\biggr].
\ee
The exchange echo then swaps $\ket{+}$ and $\ket{-}$, which followed by another evolution time of $\tau$ gives
\be
\ket{\psi(2\tau)} = \frac{1}{\sqrt{2}}\biggl[\prod_k U^-_k U^+_k\ket{-}-\prod_k U^+_kU^-_k\ket{+}\biggr].
\ee
For the simple echo, we ask the overlap of this state with the singlet $\ket{S}$:
\be
\braket{S}{\psi(2\tau)} = \frac{1}{2}\biggl[\prod_k U^-_kU^+_k+\prod_kU^+_kU^-_k\biggr].
\ee
This is an operator over all the nuclei, for which we now trace against a maximally mixed density matrix $\rho_I = 1/(2I+1)$.
As we do not simulate initialization or measurement errors, the predicted singlet-triplet decay curve goes exactly as $1/2+\exp[-\chi(t)]/2$; the decay function is therefore given for HE as
\begin{multline}
	\exp[-\chi_{\text{HE}}(t)] =
	2\Tr{\frac{1}{4}\biggl[\prod_k U^-_k U^+_k+\prod_k U^+_k U^-_k\biggr]\rho_I^{\otimes k}
		\right.\\\left.
		\biggl[\prod_k U^-_k U^+_k+\prod_k U^+_k U^-_k\biggr]^\dag}-1
	\\
	=\prod_k\frac{\Tr{U_k^- U_k^+{U_k^-}^\dag {U_k^+}^\dag}}{2I+1},
	\label{F1tau}
\end{multline}
where $t=2\tau$.
For a CP$n$ sequence, which contains $n$ $\pi$ pulses where $n$ is even, the primitive described above is simply repeated, yielding
\begin{multline}
	\exp[-\chi_{\text{CPn}}(t)]
	=
	\\
	\prod_k\frac{\Tr{[U_k^+U_k^-U_k^- U_k^+]^{n/2}[{U_k^-}^\dag {U_k^+}^\dag{U_k^+}^\dag{U_k^-}^\dag]^{n/2}}}{2I+1},
	\label{Fntau}
\end{multline}
where now $t=2n\tau$.

In practice, we perform simulations by constructing random crystals of nuclei where the probability of finding a \Ge\ nucleus at a given site is given by the product of the \Ge\ alloy content (30\% Ge in the SiGe barrier, 0\% in the Si quantum well, and following a single-monoatomic-layer-smeared profile between the two as shown in Fig.~\ref{fig:Dev}c) multiplied by the isotopic content (the 7.76\% natural abundance for \Ge).  
We then construct electron wavefunctions by solving the Schr\"odinger equation in the vertical dimension for the given Ge-content profile, assuming a linear potential offset with Ge-content and a value of 180~meV at 25\%.  
We find the phase of vertical valley oscillations by minimizing the overlap of the charge density with the random Ge nuclei placement.  
For the transverse wavefunction, we simply employ a Gaussian profile with 1/e diameter of $|\psi|^2$ at 30~nm, which is consistent with electrostatic calculations for our gate geometry.  
With the wavefunction calculated, we find the hyperfine shifts $A_k$ for every nearby \Ge\ nucleus; we localize the \Ge\ around each dot by using the 400 highest values of $A_k$ for \Ge\ nuclei in the random crystal.   
Choosing more nuclei at smaller values of $A_k$ has negligible effect on results. For each \Ge\ nucleus, we also draw its $\xi_k$ quadrupole-splitting value and electric field gradient angle $\theta_k$ values from random distributions; $\xi_k$ is drawn from a Lorentzian distribution of width $10^4$ rad/sec, while $\theta_k$ values are drawn randomly as $\tan^{-1}(x/y)$, where $x$ and $y$ are both drawn from zero-mean normal distributions with unity variance.   Once each nucleus has its parameters randomly drawn as described, then at the values of $\tau$ and $B_0^z$ of interest, we diagonalize $H^\pm_k$ for all 400 nuclei $k$ and both signs $\pm$.  
We then evaluate $\chi(t)$ according to Eqs.~(\ref{F1tau}-\ref{Fntau}).   
We then plot $1/2+\exp[-\chi(t)]/2$, as in Fig.~\ref{fig:CP10vBsim}(a), or subject this function to the same analyses as the measured singlet probabilities as in the experiments, either for spectral inversion [Eq.~(\ref{inversion_definition})] as in Fig.~\ref{fig:CP10vBsim}(b), or for finding its $1/e$ point, the $t$ such that $\chi(t)=1$, to compare to $T_2$ as in Fig.~\ref{fig:T2vWidthExp}.  

\bibliography{magnetic_noise_bibliography}

\end{document}